\documentclass[intlimits,twoside,a4paper]{article}

\usepackage{amsmath,amssymb}
\usepackage{graphicx}
\usepackage{wrapfig}

\usepackage[T2A]{fontenc}
\usepackage[cp1251]{inputenc}

\usepackage{cmpj2}

\issue{2012}{15}{2}{23801}

\doinumber{10.5488/CMP.15.23801}

\title[Contact values of profiles using density functional theory]
{On the contact values of the density profiles \\ in an electric double layer
using density \\ functional theory\thanks{It is a pleasure to dedicate
this paper to Dr. Orest Pizio on the occasion of his 60th
Birthday.  Douglas Henderson recalls with fondness his first meeting
with ``Don Oresto'' in Telavi in the Republic of Georgia in the mid 1980s,
where Orest and Myroslav Holovko invited him to visit Lviv.  During this
visit, Orest showed him the city sights, including Stefan Banach's grave
and the Scottish Cafe, where Banach and his colleagues formulated many
important theorems in functional analysis.}}

\author[L.B. Bhuiyan, D. Henderson, S. Soko\l owski]{L.B. Bhuiyan\refaddr{label1}, D. Henderson\refaddr{label2},
 S. Soko\l owski\refaddr{label3}}

\addresses{
\addr{label1}Laboratory of Theoretical Physics, Department of
Physics, University of Puerto Rico, \\ Box 70377, San Juan,  Puerto Rico 00936-8377, USA
\addr{label2}Department of Chemistry and Biochemistry, Brigham Young University,
Provo,  Utah 84602-5700, USA
\addr{label3}Department for the Modelling of Physico-Chemical Processes, Faculty of
Chemistry, MCSU, \\ 20031 Lublin, Poland
}

\date{Received September 18, 2011, in final form October 18, 2011}

\authorcopyright{L.B. Bhuiyan, D. Henderson, S. Soko\l owski, 2012}

\begin{document}

\maketitle

\begin{abstract}

A recently proposed local second contact value theorem [Henderson D.,
Boda D., J. Electroanal. Chem., 2005, \textbf{582}, 16] for the charge profile of an electric
double layer is used in conjunction with the existing Monte Carlo data from
the literature to assess the contact behavior of the electrode-ion distributions
predicted by the density functional theory. The results for the contact values
of the co- and counterion distributions and their product are obtained for the symmetric valency,
restricted primitive model planar double layer for a range of electrolyte concentrations
and temperatures. Overall, the theoretical results satisfy the second contact value theorem
reasonably well, the agreement with the simulations being semi-quantitative or better. The product
of the co- and counterion contact values as a function of the electrode surface charge density
is qualitative with the simulations with increasing deviations at higher concentrations.
\pacs 82.45.Fk, 61.20.Qg, 82.45.Gj
\end{abstract}

\section{Introduction}

One of the more interesting recent developments in the electric double
layer research has been the advancement of contact value theorems involving
the charge profile in a primitive model (PM) planar double layer (charged hard spheres
moving in a dielectric continuum next to a planar electrode) (see, for example,
references~\cite{holovko1,holdicap, hendboda,hendbhui1}). Such exact conditions, or sum rules
as they are often called, are important \emph{per se} in statistical mechanics
since they permit unambiguous assessment of various approximate theories and hence
aid in theoretical development.

    The most famous contact theorem in the double layer literature is the one
formulated by Henderson and Blum~\cite{hb}, and Henderson, Blum, and Lebowitz
(HBL)~\cite{hbl} over thirty years ago. It is a condition on the contact value
of the total density profile in a planar double layer, and for a symmetric
valency restricted primitive model (RPM) (equisized ions in the PM) planar double layer --
the model system of interest in this paper, the HBL relation reads
\begin{equation}
\label{eq:1}
g_{\rm sum}(d/2)=[g_{\rm co}(d/2)+g_{\rm ctr}(d/2)]/2=a+\frac{b^{2}}{2}\,.
\end{equation}
Here $g_{\rm co}$, $g_{\rm ctr}$ are the co- and counterion singlet distribution
functions, $d$ is the common ionic diameter, and $a=p/(\rho k_{\rm B}T)$ is the bulk
osmotic coefficient with $p$ being the bulk pressure, $k_{\rm B}$ the Boltzmann constant,
and $T$ the absolute temperature. The quantity $b=ze\sigma /(\epsilon _{0}
\epsilon _{\rm r}k_{\rm B}T\varkappa )$ is a dimensionless parameter where $z$, $e$, and $\sigma $
are, respectively, the absolute value of the ionic valency, the magnitude
of the elementary charge, and the uniform
surface charge density on the electrode, $\epsilon _{0}$ is the vacuum permittivity, and
$\epsilon _{\rm r}$ is the relative permittivity of the continuum solvent. Also,
$\varkappa =\sqrt{z^{2}e^{2}\rho /(\epsilon _{0}\epsilon _{\rm r}k_{\rm B}T)}$ is the Debye-H\"{u}ckel
parameter (inverse Debye screening length) with $\rho =\sum _{i}\rho _{i}$
where $\rho _{i}$ is the mean number density
of the $i$th ionic species. In the RPM case the contact distance, that is, the distance
of closest approach of an ion to the electrode, occurs at $d/2$, where $d_{\rm co}=d_{\rm ctr}=d$. Note that $g_{\rm sum}(x)$ ($x$ is the perpendicular distance
from the electrode into the solution) is related to the total density profile
\begin{equation}\label{eq:2}
\rho (x) =\sum_{i}\rho _{i}(x) = \sum_{i}\rho _{i}g_{i}(x) = \rho g_{\rm sum}(x),
\end{equation}
where $\rho _{i}(x)$ is the singlet density profile of the $i$th species. It is of
interest that the second term in equation~(\ref{eq:1}) is just the Maxwell stress.  Although equation~(\ref{eq:1}) was obtained from statistical mechanics it is consistent with Maxwell's equations.

Equation~(\ref{eq:1}) is a local expression and the consequent ease of its use has made
the HBL contact condition very appealing in double layer research over the years.
For example, the classical Gouy-Chapman-Stern (GCS)~\cite{gouy,chapman,stern} theory
of the double layer satisfies equation~(\ref{eq:1}) but with $a =$ 1, the ideal gas value.
Thus, for an electrolyte with osmotic coefficient substantially different from unity, the
GCS theory can lead to appreciable errors especially at low surface charges.

Sum rules such as equation~(\ref{eq:1}) are useful not only for assessing theories but also
for the insight they provide. For example, because the coion contact value becomes
small at a large surface charge and the counterion contact value becomes large,
according to this equation, the latter contact value increases as square of the surface
charge density. On the other hand, the local electroneutrality condition (also a sum-rule)
requires that the area of the charge profile be equal but opposite in sign to the electrode
charge density. As a result this area increases linearly with the electrode charge density.
Consequently, at large electrode charge, the oscillations and charge inversions in the charge profile should diminish but cannot disappear altogether relative to the contact value.

Analogous relations for the contact value of the total charge profile in the
double layer~-- the theme of the present paper, have been relatively recent. A formal,
rigorous relation was derived by Holovko et al.~\cite{holovko1,holovko2} and Holovko and di Caprio~\cite{holdicap} using the Bogoliuobov-Born-Green-Yvon hierarchy. For symmetric valency RPM planar double layer their expression is as follows:
\begin{equation}
g_{\rm diff}(d/2)=-ze\beta \int _{d/2}^{\infty}\rd x g_{\rm sum}(x)\frac{\rd\psi (x)}{\rd x}\,,
\label{eq:gdiff}
\end{equation}
where $\beta = 1/(k_{\rm B}T)$, $\psi (x)$ is the mean electrostatic potential,
and $g_{\rm diff}(x) = [g_{\rm ctr}(x)-g_{\rm co}(x)]/2$. Again, $g_{\rm diff}(x)$ is now related
to the total charge profile, viz.,
\begin{equation}
q(x)=e\sum _{i}z_{i}\rho _{i}g_{i}(x)=ze\rho g_{\rm diff}(x),
\label{eq:qx}
\end{equation}
with $z_{i}$ being the valency of the ionic species $i$ and $z = z_{\rm co} = -z_{\rm ctr}$.
The definition of $g_{\rm diff}(x)$ is, for convenience only, designed to make this quantity
positive in general. Since the use of this equation implies a knowledge of $\psi (x)$, and
$g_{i}(x)$ throughout the double layer, the expression is non-local.

Independently, Henderson and Boda (HB)~\cite{hendboda} have proposed an approximate, local
expression for $g_{\rm diff}(d/2)$ at low electrode charges from empirical considerations, viz.,
\begin{equation}
g_{\rm diff}(d/2)=ab+O(b^{3}).
\label{eq:gdiff2}
\end{equation}
To date a formal, analytic connection between equations~(\ref{eq:gdiff}) and~(\ref{eq:gdiff2}) remains
obscure, although in a later paper  Holovko et al.~\cite{holovko3} have outlined
a very approximate connection. In a series of papers Henderson and Bhuiyan~\cite{hendbhui1}
and Bhuiyan and co-workers~\cite{bhuiyan1, bhuiyan2, bhuiyan3, bhuiyan4} have tested equation~(\ref{eq:gdiff}) against \emph{exact} Monte Carlo (MC) simulation data for a spectrum of physical states including asymmetric electrolytes~\cite{bhuiyan2, bhuiyan3, bhuiyan4} and found the equation to be remarkably consistent with the simulations. Theoretical support came from an application of the modified Poisson-Boltzmann (MPB) equation, which was found to satisfy equation~(\ref{eq:gdiff}) to a very good degree~\cite{bhuiyan1,bhuiyan2}. Bhuiyan and Henderson~\cite{bhuihend} have also compared the two relations (equations~(\ref{eq:2}) and~(\ref{eq:gdiff})) numerically using simulations and the conclusion was that although exact, equation~(\ref{eq:2}) is difficult to implement numerically because of its non-local nature. We note here that an approximate, non-local relation for $g_{\rm diff}(d/2)$ has also been suggested by Henderson and Bhuiyan~\cite{hendbhui2}. Following convention, we will call equation~(\ref{eq:1}) the first contact value theorem, while equations~(\ref{eq:gdiff}) and~(\ref{eq:gdiff2})
represent two versions of the second contact value theorem.

Another interesting recent result that also concerns us in this paper is
the behavior of the product of the co- and counterion contact values
$f = g_{\rm co}(d/2)g_{\rm ctr}(d/2)$ in the RPM planar double layer. The classical GCS
result for this quantity is strictly unity under all circumstances and thus
constitutes a basis for the classical theory. For example, the value of the counterion
contact $g_{\rm ctr}(d/2)$ is as high as the reciprocal of the coion contact $g_{\rm co}(d/2)$. However, the corresponding simulation results~\cite{bhuoutdoug} dramatically alter the classical picture. The product $f$ is seen to be not
only different from unity, but also that its characteristics as a function of the electrode charge change with the salt concentration. At low concentrations there is a maximum before
$f$ becomes vanishingly small at high electrode charge. As the concentration increases, the
height of the maximum decreases and at sufficiently high concentrations the maximum
disappears completely with $f$ decreasing monotonously. Again, theoretical support for
such a behavior of $f$ came from the MPB~\cite{bhuoutdoug} and although the
hypernetted chain/mean spherical approximation theory does not show a maximum, the product
$f$ does become very small when the electrode charge is large~\cite{hasl}.

In this study we propose to utilize the HB second contact value theorem and the
existing MC simulation results from the literature for $f$ to assess the density functional
theory (DFT) of the planar double layer. The DFT has been one of the more successful
theories of the electric double layer phenomenon and compare{\bf s} favorably with the MPB
across planar, cylindrical, and spherical geometries (see for example,  references~\cite{bhuiouth1, pb, bhuiouth2}). Early applications of the DFT to the planar double layer were made by Tang et al.~\cite{tang} and Mier y Teran et al.~\cite{mieryteran1}. Later Rosenfeld's~\cite{rosenfeld} techniques were utilized by Mier y Teran et al.~\cite{mieryteran2} and Boda et al.~\cite{boda1,boda2,boda3,boda4}. For even  recent publications on application of the DFT to the planar double layer, we refer the interested reader to the works by Gillespie et al.~\cite{gillespie1, gillespie2}, Valisk\'{o} et al.~\cite{valisko}, Wang et al.~\cite{wang}, Yu et al.~\cite{yu}, and Pizio et al.~\cite{pizio}. Since there is more than one version of the DFT for the planar double layer, in the next section we will briefly outline the DFT method used in this paper. Results will be shown in section~3, and some conclusions drawn in section~4.

\section{Model and methods}

\subsection{Molecular model}

As indicated in the previous section, the model double layer system
consists of a binary, symmetric valency RPM next to a non-penetrable, non-polarizable,
uniformly charged planar electrode with a surface charge density of $\sigma $.
Since for a given salt concentration, solvent dielectric constant, and temperature,
$b$ has a linear dependence on $\sigma $, it is often convenient to specify $\sigma $
in terms of $b$.

The ion-ion interaction potential in the Hamiltonian is thus
\begin{equation}
 u_{ij}(r)=\left\{
\begin{array}{cc}
 \infty & \quad r<d, \\
e^{2} z_i z_j /(4\pi \epsilon _{0}\epsilon _{\rm r}r) & \quad r>d,
\end{array}
\right.
\end{equation}
where $r$ is the distance between a pair of ions.
We also assume that the dielectric constant, $\epsilon _{\rm r}$,  is uniform throughout the
entire system.
The bare interaction between an ion of species $i$
and the wall is given by
\begin{equation}
         u_i(x)=v_i(x)+w_i(x),
\end{equation}
where $v_i(x)$ and $w_i(x)$ are the non-electrostatic and electrostatic (Coulombic)
parts of the ion-wall potential. The non-electrostatic
contribution  is a hard-wall potential
\begin{equation}
 u_{i}(x)=\left\{
\begin{array}{cc}
 \infty & \quad x<d/2, \\
0 & \quad x>d/2.
\end{array}
\right.
\end{equation}
The electrostatic part $w_i(x)$ is given by
\begin{equation}
w_i(x)=-\frac{\sigma z_i e}{\epsilon _{0}\epsilon _{\rm r}}x,  \quad  > \frac{d}{2} \,.
\end{equation}

\subsection{Density functional theory}

The essence of the density functional theory (see for example, reference~\cite{evans1}) is that an expression for the grand potential, $\Omega$, as a functional of
the singlet density profiles, $\rho_{i}(x)$, of each of
the species $i$, is initially constructed.  At equilibrium the grand potential is
minimal with respect to variations in the density profiles, viz.,
\begin{equation}
\frac{\delta \Omega}{\delta \rho_{i}(x)}  = 0.
\label{grand}
\end{equation}
This condition is then used to calculate the density profiles and other relevant
quantities like the free energy.

In the density functional theory the grand potential of an
inhomogeneous fluid can be written in the form~\cite{boda4,pizio}
\begin{equation}
\Omega = F(\{\rho_i\})+\frac{1}{2}\sum_{i=\rm co,ctr}ez_i \int\rho_i(x)\psi(x)\rd{\bf r}+
\sum_{i= \rm co,\rm ctr}\int[u_i(x)-\mu_i]\rd{\bf r},
\label{eq:grand}
\end{equation}
where $\mu_i$ denotes the chemical potential of species $i$.
The free energy functions  $ F(\{\rho_i\})$ is decomposed into ideal ({id}),
hard-sphere ({hs}), and electrostatic ({el}) terms as follows
$ F(\{\rho_i\})= F_{\rm id}(\{\rho_i\})+ F_{\rm hs}(\{\rho_i\})+ F_{\rm el}(\{\rho_i\})$. The ideal term is known exactly
\begin{equation}
F_{\rm id}(\{\rho_i\})=\sum_{i=\rm co,ctr}\int \rd{\bf r}[\rho_i(x)\ln \rho_i(x) -\rho_i(x)].
\end{equation}
For the hard-sphere term, however, we apply the expression
resulting from a recent version of the Fundamental Measure Theory~\cite{yuwu},
with the free energy consisting of the terms dependent on scalar and vector
weighted densities, for details see reference~\cite{pizio}.

Following Pizio et al.~\cite{pizio} electrostatic contribution to the free energy, $F_{\rm el}(\{\rho_i\})$, is represented by
\begin{equation}
 F_{\rm el}(\{\rho_i\})=\int \rd{\bf r} f_{\rm el}(\{\bar \rho_i(x)\}),
\end{equation}
where $\{ \bar \rho_i (x)\}$ denotes a set of  suitably defined inhomogeneous average
densities of a reference fluid. One of the simplest possible
choices of $f_{\rm el}(\{\bar \rho_i(x)\})$  is to apply the expression resulting from
the MSA equation of state evaluated via the energy
route, namely~\cite{jbbps}
\begin{equation}
f_{\rm el}(\{\bar \rho_i(x)\})/kT=-\frac{d}{T^*}\sum_{i=\rm co,ctr}z_i^2\bar\rho_i(x)\frac{\Gamma}{1+\Gamma d}+
\frac{\Gamma^3}{3\pi}\,.
\end{equation}
For a symmetric valency situation as in the present case the reduced temperature is
$T^*=\frac{4\pi k_{\rm B}T\epsilon _ {0}\epsilon _{\rm r} d}{e^2 z^2}$
Moreover,
\begin{equation}
\Gamma=\left(\sqrt{1+2\varkappa d}-1\right)/2d.
\label{eq:gamma}
\end{equation}
The inverse Debye screening length  $\varkappa$ can be cast in terms of $T^{*}$
\begin{equation}
 \varkappa^2=(4\pi d/T^*)\sum_{i}z_i^2\bar \rho_i(x).
\end{equation}
The last three expressions above correspond
to {\it an electroneutral fluid}, so that the construction of the averaged
densities $\bar \rho_i(x)$ at the electroneutrality condition is
satisfied.  In our approach  we follow the development proposed by
Gillespie et al.~\cite{gillespie1,gillespie2} described briefly below.

Let us define the weighted densities
$\tilde \rho_i(x)$  as
\begin{equation}
 \tilde \rho_i(x)=\int \rho_i(x')W(|{\bf r}-{\bf r}'  |) \rd\bf{r}',
\label{add:1}
\end{equation}
where $W(|{\bf r}-{\bf r}'  |)$ is a weight function. Gillespie et al.
made the assumption, viz.,
\begin{equation}
W(|\mathbf{r}-\mathbf{r}'  |)=\frac{\theta(|\mathbf{r}-\mathbf{r}'  |)-R_f(\mathbf{r}')}{(4\pi/3)R_f^3(\mathbf{r}')}\,,
\label{add:2}
\end{equation}
where  $\theta(|\mathbf{r}-\mathbf{r}'  |)$ is the step-function.
The radius of the sphere over which averaging is performed,
$R_f$, is approximated by the ``capacitance'' radius, that is, by the ion radius plus
the screening length
\begin{equation}
R_f({\bf r})=\frac{d}{2}+\frac{1}{2\Gamma(\{ \bar \rho_i(x) \})}\,.
\label{add:3}
\end{equation}
In addition, Gillespie et al.~\cite{gillespie1,gillespie2} required
that the fluid with the densities $\{\bar \rho_i(x)\}$ have the same ionic
strength as the system with weighted densities,
$\{ \tilde \rho_i(x) \}$. Consequently, in the case of a
symmetric 1:1 electrolyte the averaged densities $\{ \bar \rho_i(x) \}$ are given by
\begin{equation}
\bar \rho_1(x)= \bar \rho_2(x) =\frac{\tilde \rho_1(x)+\tilde \rho_2(x)}{2}\,.
\label{add:4}
\end{equation}
Because equations (\ref{eq:gamma}) and (\ref{add:1})--(\ref{add:4}) are coupled, the evaluation of
$R_f$ requires an iteration procedure. This iteration loop has to
be carried out in addition to the main iteration procedure for
evaluating the density profiles.

The mean electrostatic potential $\psi(x)$ is determined by the Poisson
equation
\begin{equation}
 \frac {\rd^{2}\psi(x)}{\rd x^{2}}=-\frac{e}{\epsilon _{0}\epsilon _{\rm r}}\sum_{i}z_i\rho_i(x).
\end{equation}
The integration of the Poisson equation is carried out subject to the
boundary conditions $\lim_{z\to \infty} \psi(x)=0$ and  $\lim_{x\to \infty} \psi'(x)=0$.

Having specified all the contributions to the free energy functional, the requisite
density profiles can be obtained by minimizing the grand potential (cf. equation~(\ref{eq:grand})).

All the details of our approach can be found in reference~\cite{pizio}.

\section{Results and discussion}

The DFT equations have been solved numerically using
the established methods (see for example, references~\cite{tang,pizio,pg1}.
We will also present the classical GCS results for comparison purposes, which for the RPM
case can be obtained analytically. It is convenient to discuss
the results in terms of universal reduced parameters such as the reduced
density $\rho ^{*} = \sum _{i}\rho _{i}d_{i}^{3}$ and the reduced
temperature $T^{*}$ defined earlier. Calculations were done at two
different reduced temperatures, $T^{*}$ = 0.150 and 0.595, respectively, and
at each reduced temperature a number of physical states were treated. The value
of the ionic diameter was kept at $d$ = 4.25 $\times $10$^{-10}$~m throughout.
Although a 1:1 valency system was used in the actual calculations, in view
of universality of $T^{*}$ this becomes a moot point since for a given $T^{*}$
a 1:1 system at $T$ is equivalent to a 2:2 system at 4$T$. For example, in the
specific case of $T^{*}$ = 0.15, a 1:1 valency case corresponds to $\sim $75~K, while
a 2:2 valency case corresponds to $\sim $300~K.

\begin{figure}[!t]
\vspace{-1.7cm}
\includegraphics[width=0.5\textwidth]{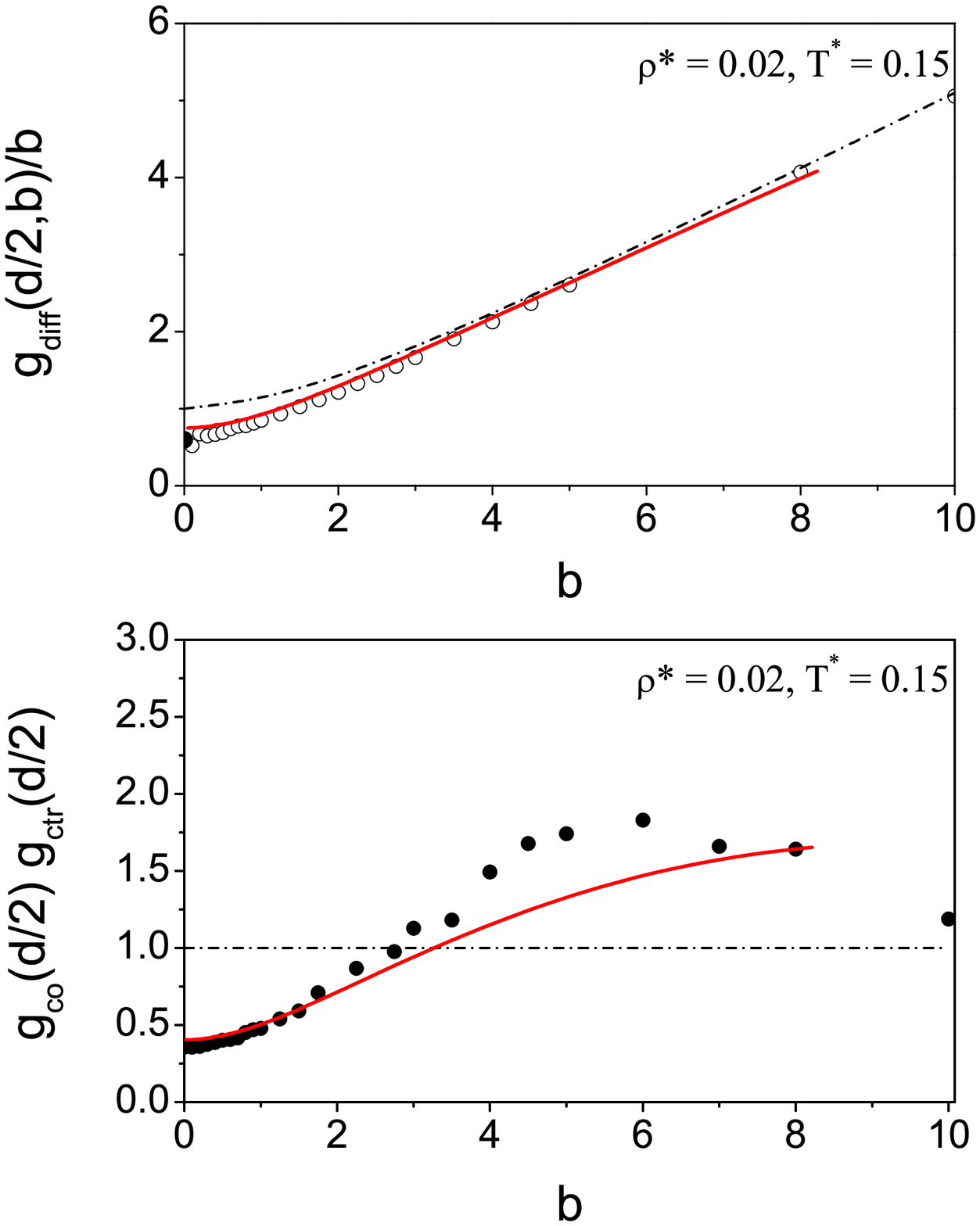}%
\hfill%
\includegraphics[width=0.5\textwidth]{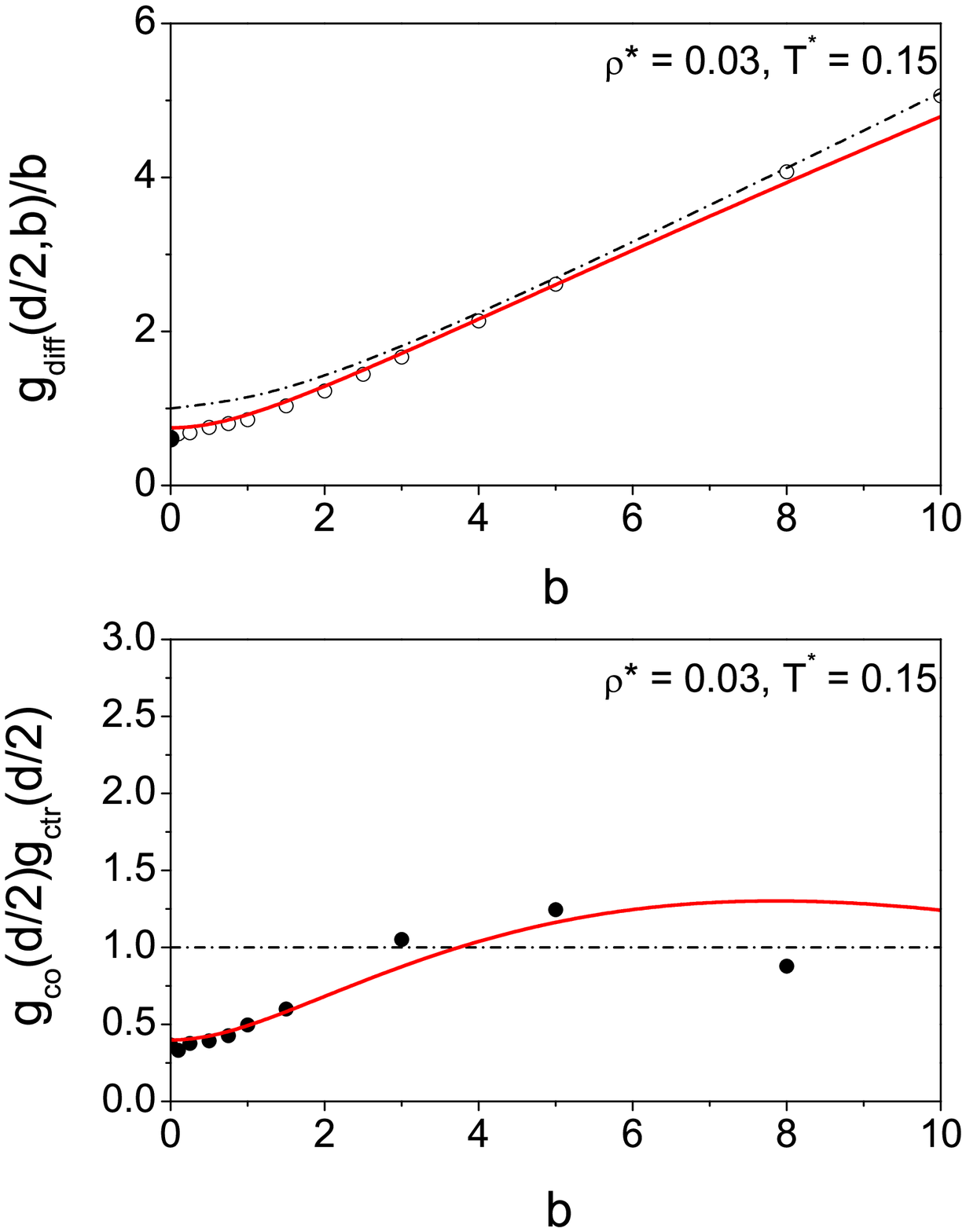}%
\\[-1.5cm]
\parbox[t]{0.5\textwidth}{%
\caption{$g_{\rm diff}(d/2,b)/b$ (upper panel) and $g_{\rm co}(d/2)g_{\rm ctr}(d/2)$
(lower panel) as functions of $b$ in a RPM planar double layer
for symmetric valencies at $\rho ^{*}$ = 0.02 ($c$ = 0.216 mol/dm$^{3}$)
and $T^{*}$ = 0.15. The symbols represent MC data, while the solid
line represents the DFT results, and the dash-dotted line the GCS results.
The filled circle on the vertical axis in the upper panel is
$g_{\rm sum}(d/2,b=0) = a$ = 0.597. MC data from reference~\cite{bhuoutdoug}.}
\label{fig:1}
}%
\hfill%
\parbox[t]{0.5\textwidth}{%
\caption{$g_{\rm diff}(d/2,b)/b$ (upper panel) and $g_{\rm co}(d/2)g_{\rm ctr}(d/2)$
(lower panel) as functions of $b$ in a RPM planar double layer
for symmetric valencies at $\rho ^{*}$ = 0.03 ($c$ = 0.324 mol/dm$^{3}$)
and $T^{*}$ = 0.15. The filled circle on the vertical axis in the upper panel is
$g_{\rm sum}(d/2,b=0) = a$ = 0.606. The rest of symbols and notation as in figure~\ref{fig:1}.
MC data from reference~\cite{bhuoutdoug}.}
\label{fig:2}
}%
\end{figure}

\begin{figure}[!b]
\vspace{-1.8cm}
\includegraphics[width=0.5\textwidth]{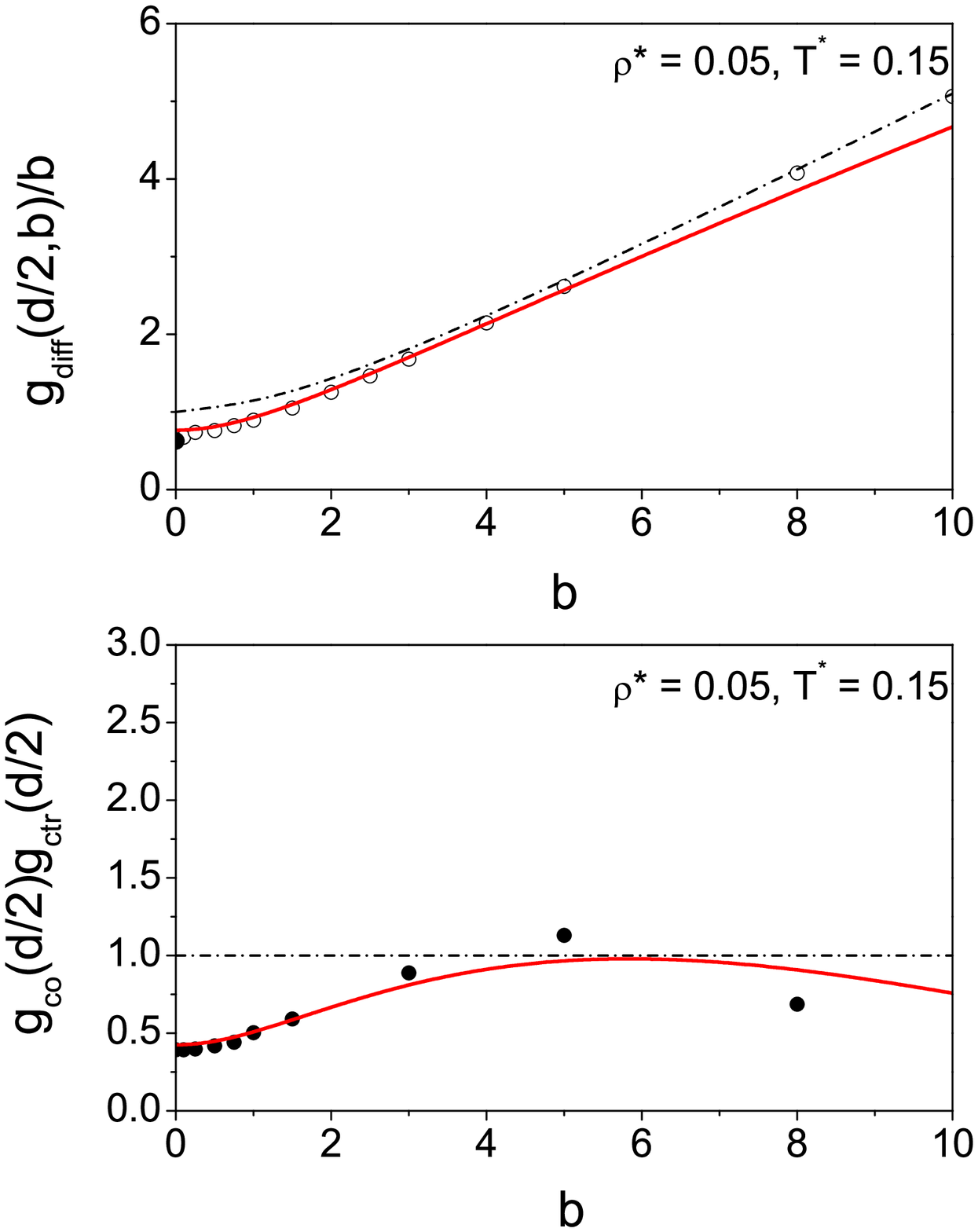}%
\hfill%
\includegraphics[width=0.5\textwidth]{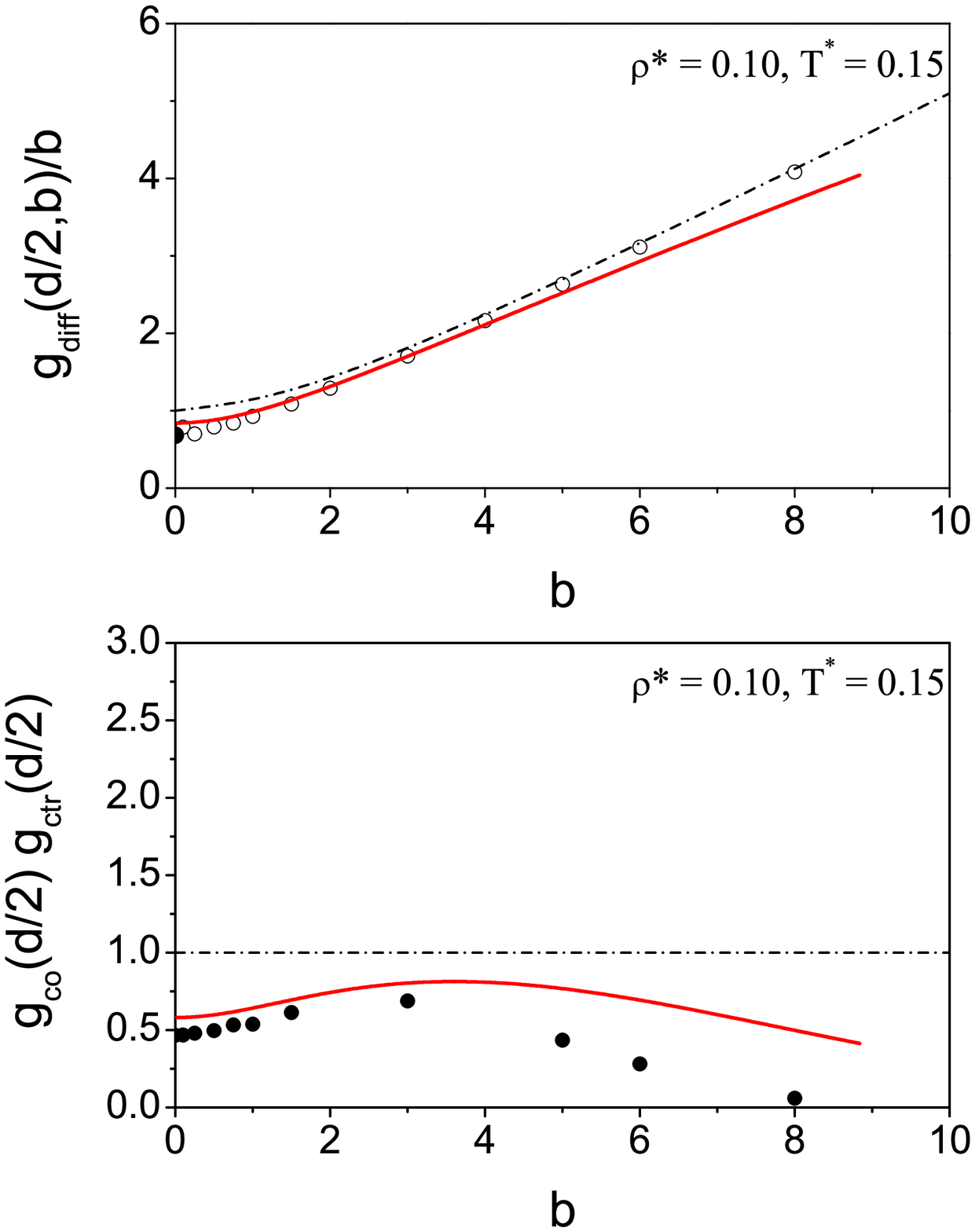}%
\\[-1.5cm]
\parbox[t]{0.5\textwidth}{%
\caption{$g_{\rm diff}(d/2,b)/b$ (upper panel) and $g_{\rm co}(d/2)g_{\rm ctr}(d/2)$
(lower panel) as functions of $b$ in a RPM planar double layer
for symmetric valencies at $\rho ^{*}$ = 0.05 ($c$ = 0.541 mol/dm$^{3}$)
and $T^{*}$ = 0.15. The filled circle on the vertical axis in the upper panel is
$g_{\rm sum}(d/2,b=0) = a$ = 0.627. The rest of symbols and notation as in figure~\ref{fig:1}.
MC data from reference~\cite{bhuoutdoug}.}
\label{fig:3}
}%
\hfill%
\parbox[t]{0.5\textwidth}{%
\caption{$g_{\rm diff}(d/2,b)/b$ (upper panel) and $g_{\rm co}(d/2)g_{\rm ctr}(d/2)$
(lower panel) as functions of $b$ in a RPM planar double layer
for symmetric valencies at $\rho ^{*}$ = 0.10 ($c$ = 1.08 mol/dm$^{3}$)
and $T^{*}$ = 0.15. The filled circle on the vertical axis in the upper panel is
$g_{\rm sum}(d/2,b=0) = a$ = 0.684. The rest of symbols and notation as in figure~\ref{fig:1}.
MC data from reference~\cite{bhuoutdoug}.}
\label{fig:4}
}%
\end{figure}

\begin{figure}[!h]
\vspace{-1.8cm}
\includegraphics[width=0.5\textwidth]{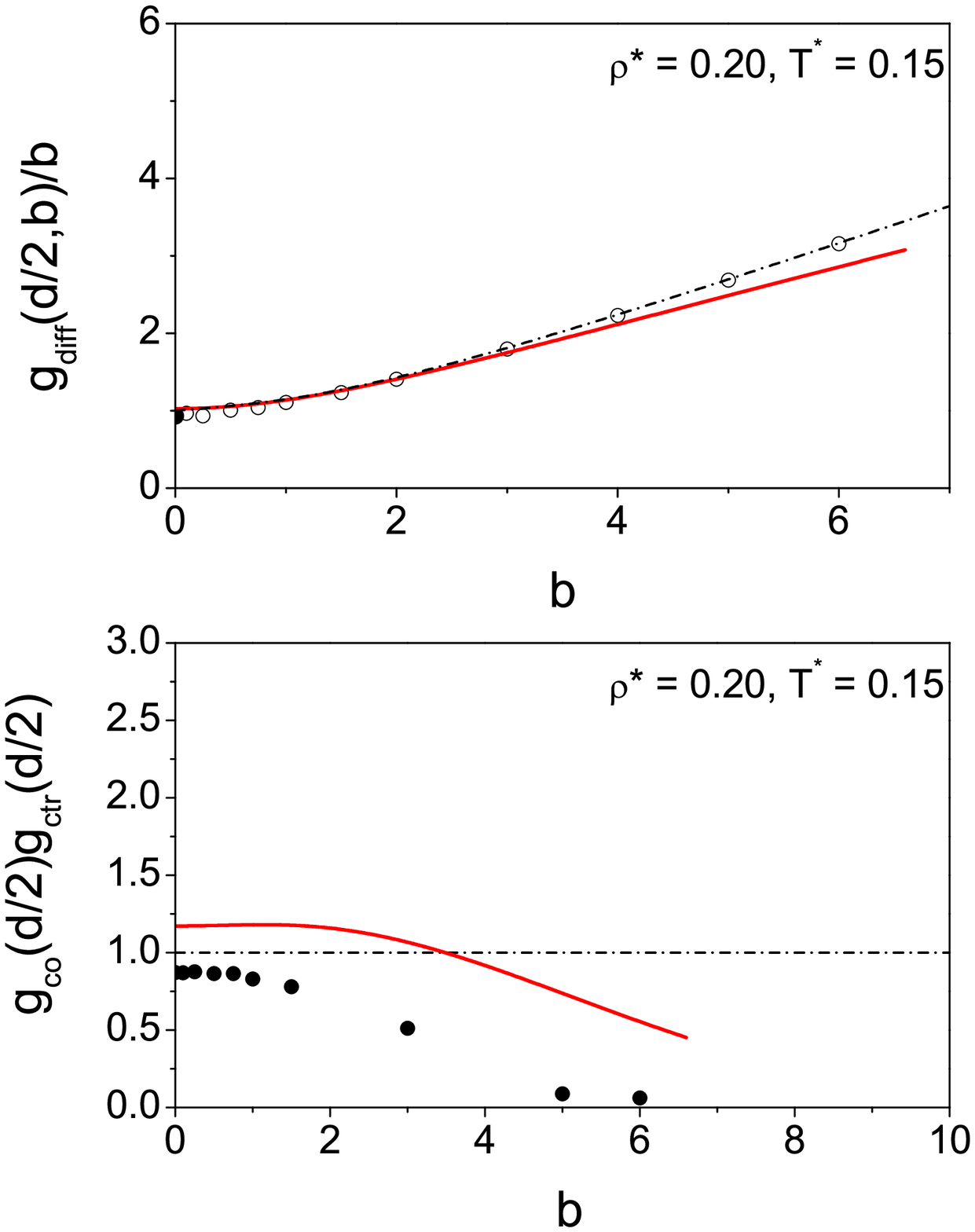}%
\hfill%
\includegraphics[width=0.5\textwidth]{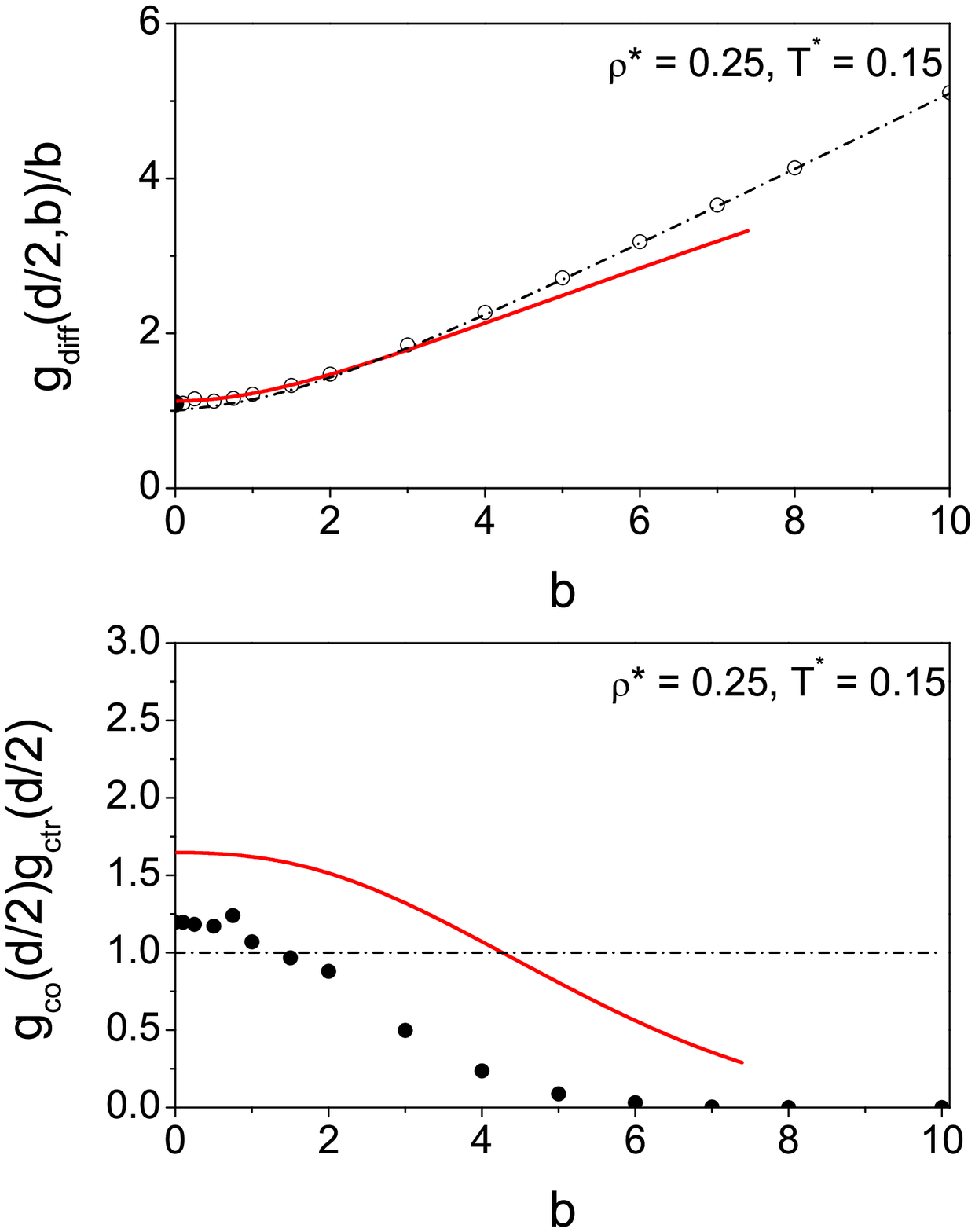}%
\\[-1.5cm]
\parbox[t]{0.5\textwidth}{%
\caption{$g_{\rm diff}(d/2,b)/b$ (upper panel) and $g_{\rm co}(d/2)g_{\rm ctr}(d/2)$
(lower panel) as functions of $b$ in a RPM planar double layer
for symmetric valencies at $\rho ^{*}$ = 0.20 ($c$ = 2.16 mol/dm$^{3}$)
and $T^{*}$ = 0.15. The filled circle on the vertical axis in the upper panel is
$g_{\rm sum}(d/2,b=0) = a$ = 0.934. The rest of symbols and notation as in figure~\ref{fig:1}.
MC data from reference~\cite{bhuoutdoug}.}
\label{fig:5}
}%
\hfill%
\parbox[t]{0.5\textwidth}{%
\caption{$g_{\rm diff}(d/2,b)/b$ (upper panel) and $g_{\rm co}(d/2)g_{\rm ctr}(d/2)$
(lower panel) as functions of $b$ in a RPM planar double layer
for symmetric valencies at $\rho ^{*}$ = 0.25 ($c$ = 2.70 mol/dm$^{3}$)
and $T^{*}$ = 0.15. The filled circle on the vertical axis in the upper panel is
$g_{\rm sum}(d/2,b=0) = a$ = 1.09. The rest of symbols and notation as in figure~\ref{fig:1}.
MC data from reference~\cite{hendbhui1}.}
\label{fig:6}
}%
\end{figure}

\begin{figure}[!h]
\vspace{-1.1cm}
\includegraphics[width=0.5\textwidth]{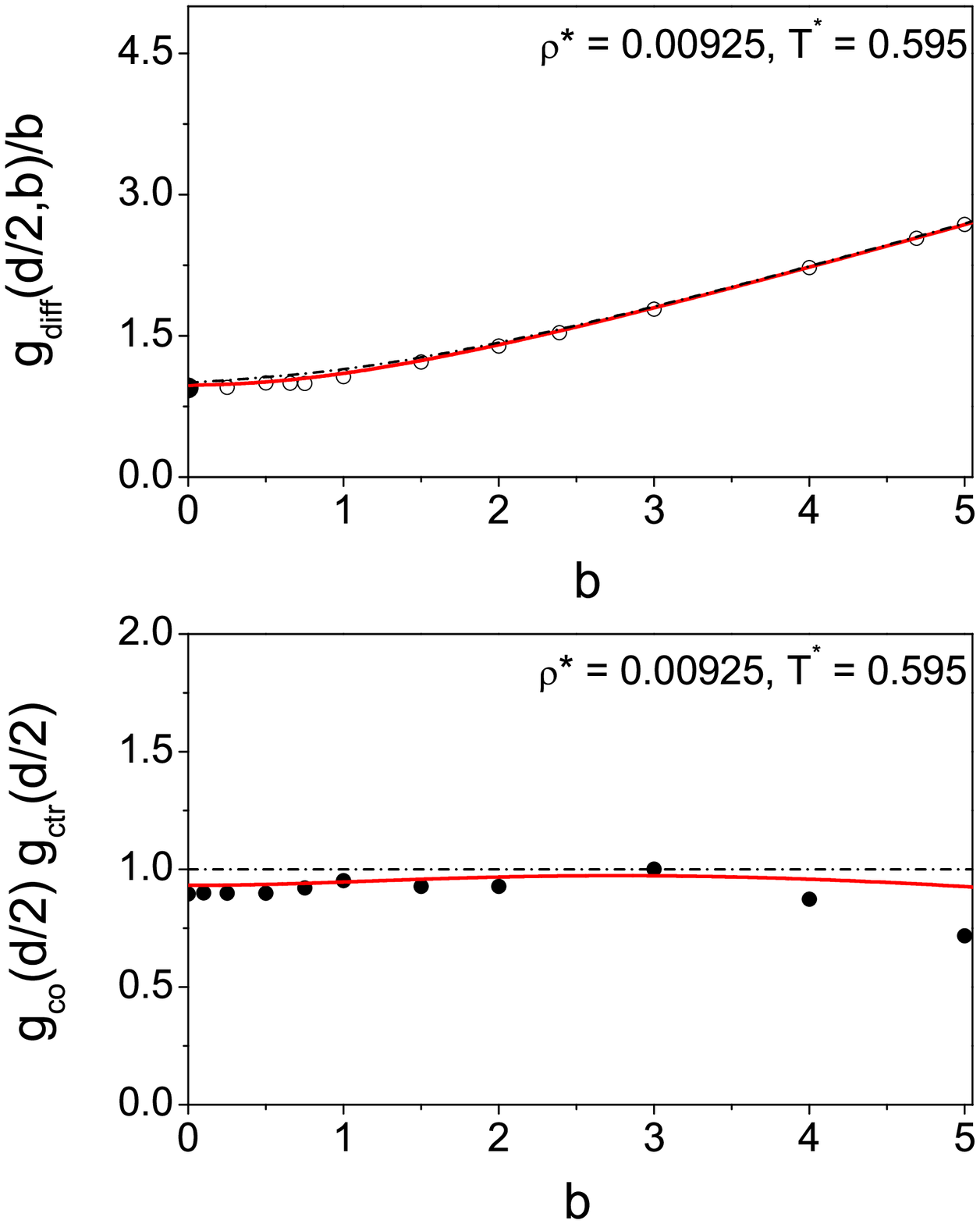}%
\hfill%
\includegraphics[width=0.5\textwidth]{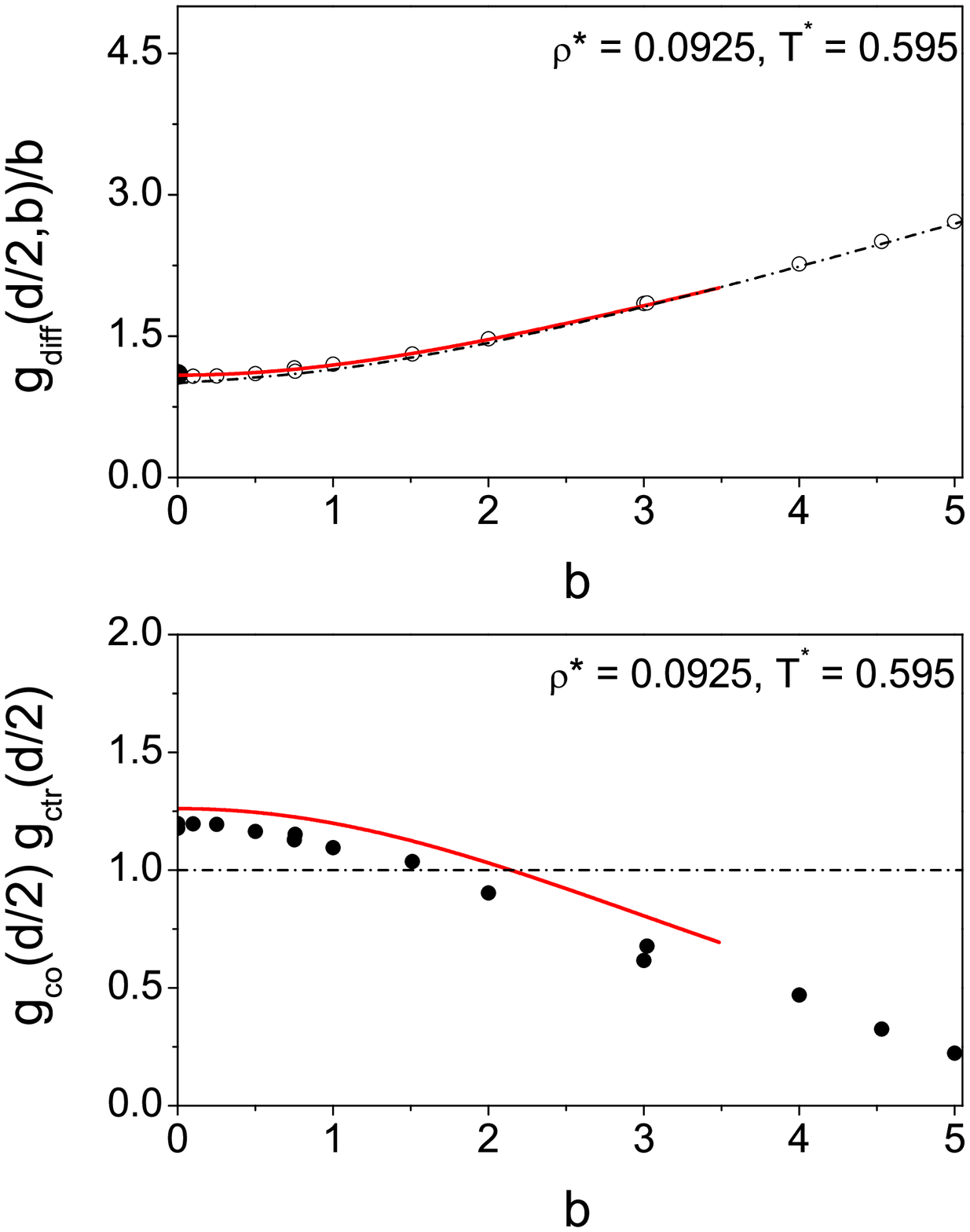}%
\\[-1.9cm]
\parbox[t]{0.5\textwidth}{%
\caption{$g_{\rm diff}(d/2,b)/b$ (upper panel) and $g_{\rm co}(d/2)g_{\rm ctr}(d/2)$
(lower panel) as functions of $b$ in a RPM planar double layer
for symmetric valencies at $\rho ^{*}$ = 0.00925 ($c$ = 0.1 mol/dm$^{3}$)
and $T^{*}$ = 0.595. The filled circle on the vertical axis in the upper panel is
$g_{\rm sum}(d/2,b=0) = a$ = 0.947. The rest of symbols and notation as in Figure 1.
MC data from reference~\cite{bhuiyan1}.}
\label{fig:7}
}%
\hfill%
\parbox[t]{0.5\textwidth}{%
\caption{$g_{\rm diff}(d/2,b)/b$ (upper panel) and $g_{\rm co}(d/2)g_{\rm ctr}(d/2)$
(lower panel) as functions of $b$ in a RPM planar double layer
for symmetric valencies at $\rho ^{*}$ = 0.0925 ($c$ = 1 mol/dm$^{3}$)
and $T^{*}$ = 0.595. The filled circle on the vertical axis in the upper panel is
$g_{\rm sum}(d/2,b=0) = a$ = 1.31. The rest of symbols and notation as in Figure 1.
MC data from reference~\cite{bhuiyan1}.}
\label{fig:8}
}%
\end{figure}

In implementing the HB contact condition, Henderson and Bhuiyan~\cite{hendbhui1}
found it convenient to recast equation~(\ref{eq:gdiff2}) in the form
\begin{equation}
\lim _{b\rightarrow 0}\left (\frac{g_{\rm diff}(d/2,b)}{b}\right )=a,
\end{equation}
for symmetrical valency electrolytes. We have followed the procedure here. We note though
that a \linebreak straightforward linear plot of equation~(\ref{eq:gdiff2}) with $a$ as the slope has
also been done~\cite{bhuihend}. In figures~\ref{fig:1}--\ref{fig:6} we present the results for
$g_{\rm diff}(d/2,b)/b$ and the contact product $f=g_{\rm co}(d/2)g_{\rm ctr}(d/2)$
as functions $b$ for $T^{*}$ = 0.15 at
$\rho ^{*}$ = 0.02 ($c$ = 0.216 mol/dm$^{3}$), 0.03 ($c$ = 0.324 mol/dm$^{3}$),
0.05 ($c$ = 0.541 mol/dm$^{3}$), 0.10 ($c$ = 1.08 mol/dm$^{3}$), 0.20 ($c$ = 2.16 mol/dm$^{3}$),
and 0.25 ($c$ = 2.70 mol/dm$^{3}$), respectively. The lone filled circle on
the vertical axis in the upper panel of a figure corresponds to the osmotic
coefficient $a$, which is evaluated from the simulations at $b$ = 0 using equation~(\ref{eq:1}).
Noticeable immediately from the figures is the trend
that the DFT $g_{\rm diff}(d/2,b)/b$ (upper panels of the figures)
follows the corresponding simulation results very closely
for not too high $b$ for the range of concentration treated. Only
a very slight discrepancy is seen at $b$ = 0, which is a consequence of the
fact that the DFT does not satisfy the HBL first contact theorem exactly.
The classical GCS theory satisfies equation~(\ref{eq:gdiff2}) but with $a$ = 1, the ideal gas value. This is clear from the figures and for $\rho ^{*}$ = 0.02, 0.03, 0.05, and 0.10
(figures~\ref{fig:1}--\ref{fig:4}), where $a$ is somewhat less than unity, the GCS theory leads to deviations
from the MC data. The results at a different temperature $T^{*}$ = 0.595 and at
$\rho ^{3}$ = 0.00925 ($c$ = 0.1 mol/dm$^{3}$) and $\rho ^{3}$ = 0.0925 ($c$ = 1.0 mol/dm$^{3}$) are shown in figures~\ref{fig:7} and~\ref{fig:8}, respectively. Here too the trends shown by the DFT $g_{\rm diff}(d/2,b)/b$
and their agreement with the corresponding simulations are similar to that seen in figures~\ref{fig:1}--\ref{fig:6}. We note that the HNC satisfies equation~(\ref{eq:1}) with the first term being a function of the hard sphere compressibility, and equation~(\ref{eq:gdiff2}) with $a=1$ in the first term.  For contact values, it is little better than the GCS theory.

The behavior of the contact product function $f$ is displayed in the
lower panel of the figures. Overall the characteristics of the DFT plots
are in qualitative agreement with the simulations. At $\rho ^{*} \leqslant $ 0.10, the MC data show a maximum with the height of the maximum decreasing as $\rho ^{*}$ increases.
The DFT result is qualitative and there is a distinct maximum
at $\rho ^{*}$ = 0.03 (figure~\ref{fig:2}), 0.05 (figure~\ref{fig:3}), and 0.10 (figure~\ref{fig:4}), and  at
$\rho ^{*}$ = 0.02, there is the hint of a maximum. Further, the maximum in the
DFT curves tends to occur at a greater value of $b$ than that in the MC. In figures~\ref{fig:5} ($\rho ^{*}$ = 0.20) and~\ref{fig:7} ($\rho ^{*}$ = 0.00925)
the plots are initially flat, while in figures~\ref{fig:6} ($\rho ^{*}$ = 0.25) and~\ref{fig:8} ($\rho ^{*}$ = 0.0925) the initial slope of $f$ is negative. In all of these figures the DFT continues to be qualitative with the simulations. The characteristics of the initial slope of $f$ as the salt concentration increases can be understood from the following. From equations~(\ref{eq:1}) and~(\ref{eq:gdiff2}) one has for low $b$
\begin{equation}
g_{\rm co}(d/2)g_{\rm ctr}(d/2) = a^{2}+(a-a^{2})b^{2},
\end{equation}
(see for example, equation (18) of reference~\cite{bhuoutdoug}). This equation is
exact in the limit $b\rightarrow $0. It is easy to see at $b$ = 0 that the value of $f$
depends on the value of $a$. Furthermore, the initial slope of $f$ is negative for
$a <$ 1 (figures~\ref{fig:1}--\ref{fig:4}), the initial slope is approximately zero and the plots are initially flat when $a \sim$ 1 (figures~\ref{fig:5} and~\ref{fig:7}), and the initial slope is negative when $a >$ 1 (figures~\ref{fig:6} and~\ref{fig:8}). Note again that since in the GCS theory $a$ = 1, the right hand side of equation~(\ref{add:3}) is unity and the initial slope is zero being consistent with the observations.

An important property of the simulation data is that the contact product $f$
tends to very small values at large values of $b$. As the surface charge increases
the coion population near the electrode is depleted, while the counterion
population increases. However, the latter also induces packing problems that
inhibit distant counterions from migrating too close to the electrode surface. All these
lead to the observed behavior of $f$. In the GCS theory however, the decrease
in $g_{\rm co}$ is always proportional to the increase in $g_{\rm ctr}$ so that
classically $f$ = 1 consistently. The DFT $f$ generally follows the MC trend in
figures~\ref{fig:2}--\ref{fig:8}.  Although in figure 1 the lack of DFT data
beyond $b$ = 8 implies that one cannot be definitive, in view of the results in the rest
of the figures, it is a fair conjecture that here also the contact product will
assume small values at still higher values of $b$.

\section{Conclusions}

In this paper we have examined the predictions of a density functional
theory of the planar electric double layer with regards to
(i) the HB second contact value theorem, and (ii) the behavior of
the product of the DFT contact values of the co- and counterion
distributions vis-a-vis exact MC simulation data from the literature.
The principal finding regarding (i) is that generally the DFT follows the
MC results very closely over the range of concentrations and temperatures
studied. There is only just a hint of discrepancy at $b$ = 0, which is
probably tied to the approximation used for the hard-sphere term in the free energy
functional used to construct the grand potential. By contrast, the GCS results
show greater deviations from the simulations, especially at lower concentrations
when the MC $a$ is less than unity. It is of interest to note that the degree to which
the DFT satisfies the HB contact condition is very similar to what some of us have
observed with the MPB theory~\cite{bhuiyan1,bhuiyan2} with both of the approaches showing
slight deviations at $b$ = 0. This is not surprising since none of the theories
satisfies the HBL first contact value theorem exactly.

With respect to (ii) above, on the other hand, our calculations indicate that the DFT is broadly in qualitative agreement with the characteristics of the simulations, with the product $f$ of the contact values tending to small
values with increasing surface charge on the electrode. A maximum in $f$ as a
function of $b$ seen at lower concentrations although this occurs as, what might be
termed, a \emph{delayed maximum }. Importantly though, the behavior of the the initial
slope of $f$ as the electrolyte concentration increases follows the MC trend.
Admittedly, however, there is a quantitative discrepancy between the DFT results and the
MC data beyond $c \sim $ 1 mol/dm$^{3}$. This is understandable in view of the fact that
the product of the contact values of the distributions can be a rather more sensitive
quantity than their difference so that a slight error in either of the contact values
tends to become magnified in the contact product.

The version of density functional that is employed in this paper gives good results for
the contact values but is less satisfactory in predicting oscillatory profiles.  In contrast, other versions of density functional theory [35,40] are better at producing oscillatory profiles but are less successful for contact values.  There is more to be done in the development of a fully satisfactory density functional theory.

\ukrainianpart

\title{
Контактні значення профілів густини в електричному подвійному шарі, використовуючи теорію\\ функціоналу густини
}
\author{Л.Б. Бгуян\refaddr{label1}, Д. Гендерсон\refaddr{label2}, С. Соколовскi\refaddr{label3} }
\addresses{
\addr{label1}Лабораторiя теоретичної фiзики, фiзичний факультет, Унiверситет Пуерто Рiко, США
\addr{label2}Факультет хiмiї i бiохiмiї, Унiверситет Брiгема Янга, Прово,  США
\addr{label3}Вiддiл моделювання фiзико-хiмiчних процесiв, хiмiчний факультет, \\
 Унiверситет iм. Марiї Складовської-Кюрi, Люблiн, Польща
}

\makeukrtitle

\begin{abstract}
\tolerance=3000%
 Нещодавно запропоновану теорему про локальне друге контактне значення
 [Henderson D., Boda D., J. Electroanal. Chem., 2005, \textbf{582}, 16] для профілю заряду електричного подвiйного шару
 поєднано з iснуючими в лiтературi даними Монте Карло з метою оцінки
 контактної поведiнки електрод-iонних розподiлiв, передбачених теорiєю
 функцiоналу густини.
 Результати для контактних значень розподiлiв ко- i протиiонiв та їхнього
 добутку отримано для випадку симетричної
 валентностi в рамках обмеженої примiтивної моделi плоского подвiйного шару для
 низки концентрацiй i
 температур електролiту. В цiлому, теоретичнi результати досить добре
 задовольняють теорему про друге
 контактне значення, узгодження із симуляцiями~-- напiвкiлькiсне або краще.
 Добуток ко- i
 протиiонних контактних значень як функцiя густини заряду поверхнi електрода
 якiсно узгоджується
 з симуляцiями, але вiдхилення між обома зростає при вищих концентрацiях.
\keywords електричний подвiйний шар, обмежена примiтивна модель, профiлi
 густини
\end{abstract}


\begin{thebibliography}{99}
%
\bibitem{holovko1} Holovko M., Badiali J., di Caprio D., J. Chem. Phys., 2005,
{\bf 123}, 234705; \doi{10.1063/1.2137707}.
%
\bibitem{holdicap} Holovko M., di Caprio D., J. Chem. Phys., 2008,
{\bf 128}, 174702; \doi{10.1063/1.2909973}.
%
\bibitem{hendboda} Henderson D., Boda D., J. Electroanal. Chem., 2005, \textbf{582}, 16;
\doi{10.1016/j.jelechem.2004.11.027}.
%
\bibitem{hendbhui1} Henderson D., Bhuiyan L.B., Mol. Simulat., 2007, {\bf 33}, 953;
\doi{10.1080/08927020701461247}.
%
\bibitem{hb} Henderson D., Blum L., J. Chem. Phys., 1978, {\bf 69}, 5441;
\doi{10.1063/1.436535}.
%
\bibitem{hbl} Henderson D., Blum L., Lebowitz J.L., J. Electroanal. Chem., 1979, {\bf 102}, 315; \\
\doi{10.1016/S0022-0728(79)80459-3}.
%
\bibitem{gouy} Gouy G., J. Phys. (Paris), 1910, {\bf 9}, 457;
\doi{10.1051/jphystap:019100090045700}.
%
\bibitem{chapman} Chapman D.L., Philos. Mag., 1913, {\bf 25}, 475;
\doi{10.1080/14786440408634187}.
%
\bibitem{stern} Stern O., Elektrochem., 1924, {\bf 30}, 508.
%
\bibitem{holovko2} Holovko M., Badiali J., di Caprio D., J. Chem. Phys., 2007, {\bf 127}, 014106;
\doi{10.1063/1.2750336}.
%
\bibitem{holovko3} Holovko M., Badiali J., di Caprio D., J. Chem. Phys., 2008, {\bf 128}, 117102;
\doi{10.1063/1.2873466}.
%
\bibitem{bhuiyan1} Bhuiyan L.B., Outhwaite C.W., Henderson D., Alawneh M., Bangladesh
J.~Phys.,
2007, {\bf 4}, 93.
%
\bibitem{bhuiyan2} Bhuiyan L.B., Outhwaite C.W., Henderson D., Mol. Phys., 2009, {\bf 107}, 343; \\ \doi{10.1080/00268970902758649}.
%
\bibitem{bhuiyan3} Bhuiyan L.B., Henderson D., Mol. Simulat., 2011, {\bf 37}, 269;
\doi{10.1080/08927022.2010.502561}.
%
\bibitem{bhuiyan4} Bhuiyan L.B., Henderson D., Mol. Phys. (in press).
%
\bibitem{bhuihend} Bhuiyan L.B., Henderson D., J. Chem. Phys., 2008, {\bf 128}, 117101;
\doi{10.1063/1.2873370}.
%
\bibitem{hendbhui2} Henderson D.J., Bhuiyan L.B., Collect. Czech. Chem. Commun., 2008, {\bf 73}, 558; \\ \doi{10.1135/cccc20080558}.
%
\bibitem{bhuoutdoug} Bhuiyan L.B., Outhwaite C.W., Henderson D., J. Electroanal. Chem., 2007, {\bf 607}, 54; \\  \doi{10.1016/j.jelechem.2006.10.010}.
%
\bibitem{hasl} Henderson D., Alawneh M., Saavedra-Barrera R., Lozada-Cassou M., Condens. Matter Phys.,
2007, {\bf 10}, 323.
%
\bibitem{bhuiouth1} Bhuiyan L.B., Outhwaite C.W., Phys. Chem. Chem. Phys., 2004, {\bf 6}, 3467;
\doi{10.1039/b316098j}.
%
\bibitem{pb} Patra C.N., Bhuiyan L.B., Condens. Matter Phys., 2005, {\bf 8}, 425.
%
\bibitem{bhuiouth2} Bhuiyan L.B., Outhwaite C.W., Condens. Matter Phys., 2005, {\bf 8}, 287.
%
\bibitem{tang} Tang Z., Mier y Teran L., Davis H.T., Scriven L.E., White H.S., Mol. Phys., 1990,
{\bf 71}, 369; \\ \doi{10.1080/00268979000101851}.
%
\bibitem{mieryteran1} Mier y Teran L., Tang Z., Davis H.T., Scriven L.E., White H.S., Mol. Phys., 1991,
{\bf 72}, 817; \\ \doi{10.1080/00268979100100581}.
%
\bibitem{rosenfeld} Rosenfeld J., J. Chem. Phys., 1993, {\bf 98}, 8126; \doi{10.1063/1.464569}.
%
\bibitem{mieryteran2} Mier y Teran L., Boda D., Henderson D., Qui\~{n}iones S., Mol. Phys., 2001, {\bf 99}, 1323; \\  \doi{10.1080/00268970110048383}.
%
\bibitem{boda1} Boda D., Henderson D., Rowley R., Soko{\l}owski S., J. Chem. Phys., 1999, {\bf 111}, 9382; \\ \doi{10.1063/1.479850}.
%
\bibitem{boda2} Boda D., Henderson D., Patrykiejew A., Soko{\l}owski S., J. Chem. Phys., 2000, {\bf 113}, 802; \\ \doi{10.1063/1.481855}.
%
\bibitem{boda3} Boda D., Fawcett W.R., Henderson D., Soko{\l}owski S.,  J. Chem. Phys., 2002, {\bf 116}, 7170; \\ \doi{10.1063/1.1464826}.
%
\bibitem{boda4}
Boda~D., Henderson~D., L. Mier y Teran, S. Soko\l owski, J.
Phys.: Condens. Matter, \textbf{14}, 11945 (2002); \\
\doi{10.1088/0953-8984/14/46/305}.
%
\bibitem{gillespie1} Gillespie D., Valisk\'{o} M., Boda D., J. Phys.: Condens. Matter, 2005, {\bf 17}, 6609; \\
\doi{10.1088/0953-8984/17/42/002}.
%
\bibitem{gillespie2} Gillespie D., Nonner W., Eisenberg R., J. Phys.: Condens. Matter, 2002 {\bf 14} 12129; \\
\doi{10.1088/0953-8984/14/46/317}.
%
\bibitem {valisko} Valisk\'{o} M., Boda D., Gillespie D., J. Phys. Chem. C, 2007, {\bf 111}, 15575;
\doi{10.1021/jp073703c}.
%
\bibitem{wang} Wang K., Yu Y.-X., Gao G.-H., J. Chem. Phys., 2005, {\bf 123}, 234904;
\doi{10.1063/1.2137710}.
%
\bibitem{yu} Yu Y.-X., Wu J., Gao G.-H., Chinese J. Chem. Eng., 2004, {\bf 12}, 688.
%
\bibitem{pizio}
Pizio O., Patrykiejew A., Soko{\l}owski S., J. Chem. Phys., 2004,
\textbf{121}, 11957; \doi{10.1063/1.1818677}.
%
\bibitem{evans1} Evans R., Fundamentals of Inhomogeneous Fluids, ed. D. Henderson. Dekker, New York, 1992.
%
\bibitem{yuwu} Yu Y.-X., Wu J.Z., J. Chem. Phys., 2002, \textbf{117}, 10156;
\doi{10.1063/1.1520530}.
%
\bibitem{jbbps}
Jiang J., Blum L., Bernard O., Prausnitz J.M., Sandler S.I., J.
Chem. Phys., 2002, \textbf{116}, 7977; \\ \doi{10.1063/1.1468638}.
%
\bibitem{pg1} Patra C.N., Ghosh S.K., Phys. Rev. E, 1993, \textbf{47}, 4088;
\doi{10.1103/PhysRevE.47.4088}.
%
\end{thebibliography}
\end{document}